\documentclass[prl,nofootinbib,showpacs,twocolumn,superscriptaddress]{revtex4}
\usepackage{graphicx}
\usepackage{bm}
\usepackage{amssymb,amsmath}
\usepackage{epstopdf}
\usepackage{epsfig,dcolumn}
\usepackage{array}

\def\beqn{\begin{eqnarray}} 
\def\eeqn{\end{eqnarray}}

\begin{document}

\title{Light-front interpretation of Proton Generalized Polarizabilities}
\author{M. Gorchtein}
\affiliation{Indiana University, Bloomington, IN 47408, USA}

\author{C. Lorc\'e}
\affiliation{Institut f\"ur Kernphysik, Mainz Universit\"at, 55099 Mainz, Germany}

\author{B. Pasquini}
\affiliation{
Dipartimento di Fisica Nucleare e Teorica, Universit\`{a} di Pavia, and 
Istituto Nazionale di Fisica Nucleare, Sezione di Pavia, I-27100 Pavia, Italy}

\author{M. Vanderhaeghen}
\affiliation{Institut f\"ur Kernphysik, Mainz Universit\"at, 55099 Mainz, Germany}

\date{\today}
\begin{abstract}
We extend the recently developed formalism to extract light-front quark charge densities from nucleon form factor data to the deformations of these quark charge densities when 
applying an external electric field. We show that the resulting induced polarizations can be 
extracted from proton generalized polarizabilities. 
The available data for the generalized electric polarizabilitiy of the proton yield a pronounced structure in its induced polarization at large transverse distances, which will be pinned down by forthcoming 
high precision virtual Compton scattering experiments.
\end{abstract}

\maketitle

The distribution of charge is a basic quantity which characterizes a many-body system. In the 
case of relativistic many-body systems such as hadrons, composed of near massless quarks, 
a field theoretic consistent charge density can be formulated by considering the system 
in a light-front frame. In such frame, the pair creation by the probing photon is suppressed, and the 
photon only couples to forward moving quarks, allowing for a density interpretation. 
Such a charge density interpretation, based on elastic form factor data,   
was recently given for the nucleon \cite{Miller:2007uy, Carlson:2007xd}, 
for spin-1 systems~\cite{Carlson:2008zc}, spin-3/2 systems~\cite{Alexandrou:2009hs}, and 
extended to higher spin systems in~\cite{Lorce:2009bs}.  

Any charge density will deform when subjected to an external electric field and develop an induced 
polarization. The quantity describing the ``ease" by which such distribution will deform is referred to as the electric polarizability. 
In this work we will extend the formalism of light-front charge densities to 
obtain the spatial deformations of these densities. We will show that the resulting induced polarizations can be obtained from nucleon generalized 
polarizabilities (GPs)~\cite{Guichon:1995pu, Drechsel:1997xv}, 
which have been measured in recent years by precision virtual Compton scattering (VCS) experiments, see Refs.~\cite{Guichon:1998xv, Drechsel:2002ar} for reviews. 

We consider the VCS process on the nucleon 
$\gamma^*(q)+N(p)\to\gamma(q')+N(p')$. Its kinematics are described in terms of 
Lorentz scalars~: $Q^2 = - q^2$, $\nu \equiv q \cdot P / M$, with $P = (p + p^\prime)/2$, 
and $t = (p - p^\prime)^2$. 
The dynamical information which is accessed in the VCS process is described by the matrix element 
of a time-ordered product of two electromagnetic (e.m.) current operators as~:
\begin{eqnarray}
H^{\mu \nu} = -i \int d^4 x \, e^{- i q \cdot x} \langle p^\prime, \lambda^\prime_N | 
T \left[ J^\mu(x), J^\nu(0) \right] | p,   \lambda_N \rangle,
\label{eq:tordered}
\end{eqnarray}
with $\lambda_N$ ($\lambda^\prime_N$) the helicities of the initial (final) nucleons. 
In this work, we consider the VCS tensor in the low-energy limit, $q^\prime \to 0$. 
In such a limit, the final soft photon plays the role of an applied quasi-static electromagnetic field, 
and the VCS process measures the linear response of the nucleon to this 
applied field~\cite{Guichon:1995pu,Guichon:1998xv}. 
This linear response can be parameterized through six $Q^2$ dependent GPs, 
denoted by $P^{(\rho' \, l', \rho \,l)S}$
~\cite{Guichon:1995pu, Drechsel:1997xv}. 
In this notation, $\rho$ ($\rho'$) refers to the
Coulomb/electric ($L$), or magnetic ($M$) nature of the initial 
(final) photon, $l$ ($l' = 1$) is the angular momentum of the
initial (final) photon, and $S$ differentiates between the 
spin-flip ($S=1$) and non spin-flip ($S=0$) transition at the nucleon side.

To arrive at a spatial representation of the information contained in the GPs, 
we consider the process in a symmetric light-front frame, denoting the average direction of the fast moving protons 
as the $z$-axis. We indicate the (large) light-front + component by $P^+$ 
(defining $a^\pm \equiv a^0 \pm a^3$), and choose the symmetric frame by requiring that 
$\Delta = p^\prime - p$ is purely transversal, i.e. $\Delta^+ = 0$. 
To access the GPs, we can restrict ourselves to the terms in the VCS tensor that are linear in the 
outgoing photon energy (proportional to $\nu$), along the line $t = -Q^2$.  
In this limit the light-front kinematics is given by~:
$P^\mu = (P^+ / 2) \, \bar n^\mu + M^2 (1 + \tau) / (2 P^+) \, n^\mu$,
$\Delta^\mu = q^\mu_\perp$, 
$q^\mu = (\eta P^+)  \, \bar n^\mu + q^\mu_\perp$, 
$q^{\prime \, \mu} = (\eta P^+) \, \bar n^\mu$,
with light-like vectors $\bar n = (1, 0, 0, 1)$,  $n = (1, 0, 0, -1)$. 
Furthermore, the two transverse components of the 
virtual photon momentum are denoted by $\vec q_\perp$ with $Q^2 = \vec q_\perp^{\, 2}$, and 
$\tau \equiv Q^2 / (4 M^2)$, with $M$ the nucleon mass.  
The (small) momentum fraction $\eta$ is obtained as $\eta = \nu / (M (1 + \tau))$. 

In the light-front frame, the + component of the current $J^\mu$ in~(\ref{eq:tordered}) 
is a positive definite operator for each quark flavor, 
allowing for a light-front charge density interpretation.    
The VCS light-front helicity amplitudes can then be obtained from the VCS tensor $H^{\mu \nu}$ as~: 
\begin{eqnarray}
 H(\lambda^\prime_\gamma, \lambda^\prime_N , \lambda_N) 
\equiv  \varepsilon^{\, \prime *}_\nu (\lambda^\prime_\gamma) \, H^{+ \nu},
\label{eq:vcshel} 
\end{eqnarray}
with transverse outgoing photon polarization vector denoted by   
$\vec \epsilon^{\, \prime}_\perp $, and $\lambda^\prime_\gamma = \pm 1$   denoting 
its helicity. In the following, we will consider the polarization component of the 
outgoing photon corresponding with an electric field, $\vec E = - \partial \vec A / \partial t$, 
which can be expressed as~: 
\begin{equation}
\vec E \sim i \, q^{\prime 0}  \, \vec \epsilon^{\, \prime}_\perp = i \, 
\frac{\nu}{(1 + \tau)} \frac{P^+}{M} \vec \epsilon^{\, \prime}_\perp . 
\label{eq:elfield} 
\end{equation}
Any system of charges will respond to such an applied electric field, 
resulting in an induced polarization $\vec P_0$, 
which will be forced to align with the applied electric field such as to minimize its energy 
$ - \vec E \cdot \vec P_0$.  
 The linear response in $q^{\prime 0}$ of the helicity averaged VCS amplitude therefore 
 allows to define an induced polarization $\vec P_0$ as~:
 \begin{equation}
 i \, \vec \varepsilon_\perp^{\, \prime *} (\lambda^\prime_\gamma) \cdot \vec P_0 
 \equiv \frac{(1 + \tau)}{(2 P^+)}
 \frac{\partial}{\partial \nu}  
  H \left(\lambda^\prime_\gamma, \lambda_N, \lambda_N \right) \big|_{\nu = 0} .
  \label{eq:indpol}
 \end{equation}
The induced polarization $\vec P_0$ for the helicity averaged case can be worked out from 
Eq.~(\ref{eq:vcshel}) as~:
\begin{eqnarray}
\vec P_0(\vec q_\perp) = i \, \hat q_\perp \,  A(Q^2) ,
\label{eq:indpolnoflip}
\end{eqnarray}
where $A$ can be expressed in terms of the GPs as~:
\begin{eqnarray}
&& \hspace{-0.3cm} A = - (2 M) \, \sqrt{\tau} \,
\sqrt{\frac{3}{2}} \sqrt{\frac{1 + 2 \tau}{1 + \tau}} \nonumber \\
&& \hspace{-0.3cm} \times  \left\{ -  P^{(L1, L1)0} + \frac{1}{2} P^{(M1, M1)0} - \sqrt{\frac{3}{2}} P^{(L1, L1)1}   \right. \nonumber \\
&&\left. \hspace{-0.3cm}
- \sqrt{\frac{3}{2}} (1 + \tau) \left[ P^{(M1, M1)1} + \sqrt{2} \, (2 M \tau) P^{(L1, M2)1} \right]  
\right\} , \quad  
\label{eq:lfnohelflip2}
\end{eqnarray}
which depends on the scalar GPs, as well as those 
spin GPs  which enter the unpolarized VCS response functions.  

In an analogous way, we can define the linear response to an external 
quasi-static e.m. dipole field when the nucleon is in an eigenstate of transverse spin, 
$\vec S_\perp \equiv \cos \phi_S \hat e_x + \sin \phi_S \hat e_y$,  
with $\phi_S$ the angle indicating the spin vector direction.  
Analogously to Eq.~(\ref{eq:indpol}), the 
induced polarization $\vec P_T$ for a state of transverse spin 
can be worked out from the sum of contributions from spin-averaged and spin-flip light-front helicity 
amplitudes as~:
\begin{eqnarray}
\vec P_T (\vec q_\perp) &=& i \, \hat q_\perp \, A(Q^2) + \hat q_\perp \, 
\left( \vec S_\perp \times \vec e_z \right) \cdot \hat q_\perp \, B(Q^2) 
\nonumber \\
&+&  \left( \vec S_\perp \times \vec e_z \right)  \, C(Q^2). 
\end{eqnarray}
The functions $B$ and $C$ entering  the induced polarization $\vec P_T$  
can be expressed in terms of the GPs as~: 
\begin{eqnarray}
B &=&  -  (2 M) \tau \; 
\sqrt{\frac{3}{2}} \sqrt{\frac{1 + 2 \tau}{1 + \tau}} 
  \left\{ P^{(L1, L1)0} - \frac{1}{2} P^{(M1, M1)0}   \right. \nonumber \\
 &+&
\sqrt{\frac{3}{2}} \left[ 
- \sqrt{2} (2 M) (1 + \tau) P^{(L1, M2)1} + P^{(L1, L1)1} \right. \nonumber \\
& +& \left.  \left. \, \sqrt{\frac{3}{2}}  (2 M)(1 + \tau) P^{(M1, L2)1} \right]  
\right \}, \\
 C &=&   (2 M) \, ( 1 + \tau) \,  \frac{3}{2}  \sqrt{\frac{1 + 2 \tau}{1 + \tau}} \nonumber \\
&& \hspace{-1cm} \times
\left\{ P^{(M1, M1)1} +  
P^{(L1, L1)1} + \sqrt{\frac{3}{2}} (2 M \tau)  P^{(M1, L2)1}  
 \right\}.
\label{eq:lfhelflip2}
\end{eqnarray}

\begin{figure}
\vspace{-1cm}
\begin{center}
\includegraphics[width = 8.cm]{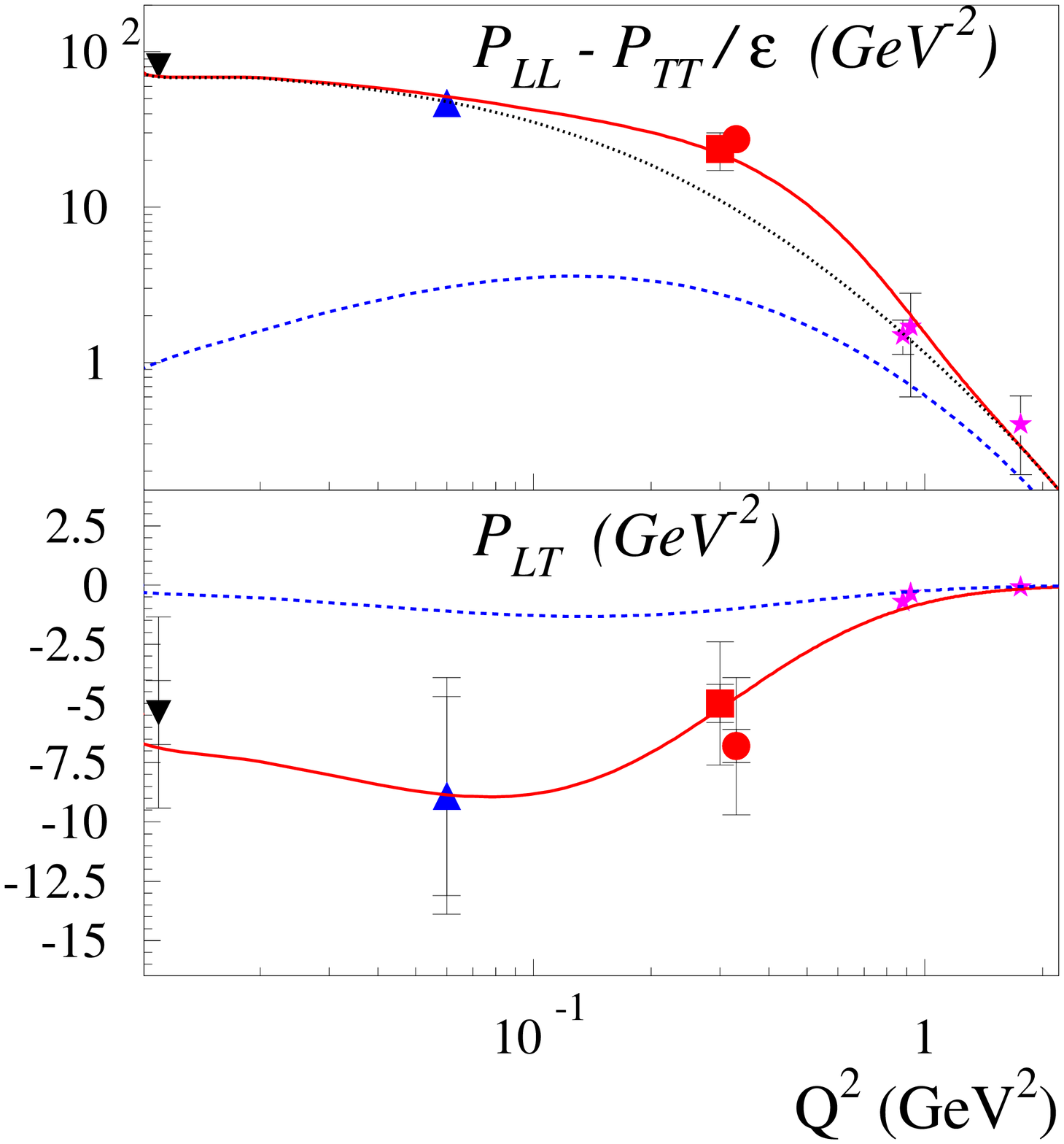}
\end{center}
\vspace{-0.5cm}
\caption{Structure functions describing unpolarized VCS on a proton compared with data from  MAMI (circles~\cite{Roche:2000ng}, squares~\cite{Janssens:2008qe}), 
MIT-Bates (up triangles~\cite{Bourgeois:2006js}), 
and JLab (stars~\cite{Laveissiere:2004nf}). 
The RCS data~\cite{OlmosdeLeon:2001zn} are shown by the (black) down triangles (slightly displaced in $Q^2$). 
The curves are based on the parameterizations of Eqs.~(\ref{eq:gpmagn}, \ref{eq:gpelec})
for the proton GPs, and are shown for $\varepsilon = 0.645$. 
Upper panel~: 
dotted (black) curve for $\Lambda_\alpha = 0.7$~GeV,  $C_\alpha = 0$ (GP I); 
solid (red) curve for $\Lambda_\alpha = 0.7$~GeV,  $C_\alpha = -150$~GeV$^{-7}$ (GP II).
Lower panel~: solid (red) curve for $\Lambda_\beta = 0.5$~GeV.  
The dashed (blue) curves in both panels show the spin GP contributions.}
\label{fig:pllplt}
\end{figure}

\begin{figure}
\begin{center}
\includegraphics[width = 6.25cm]{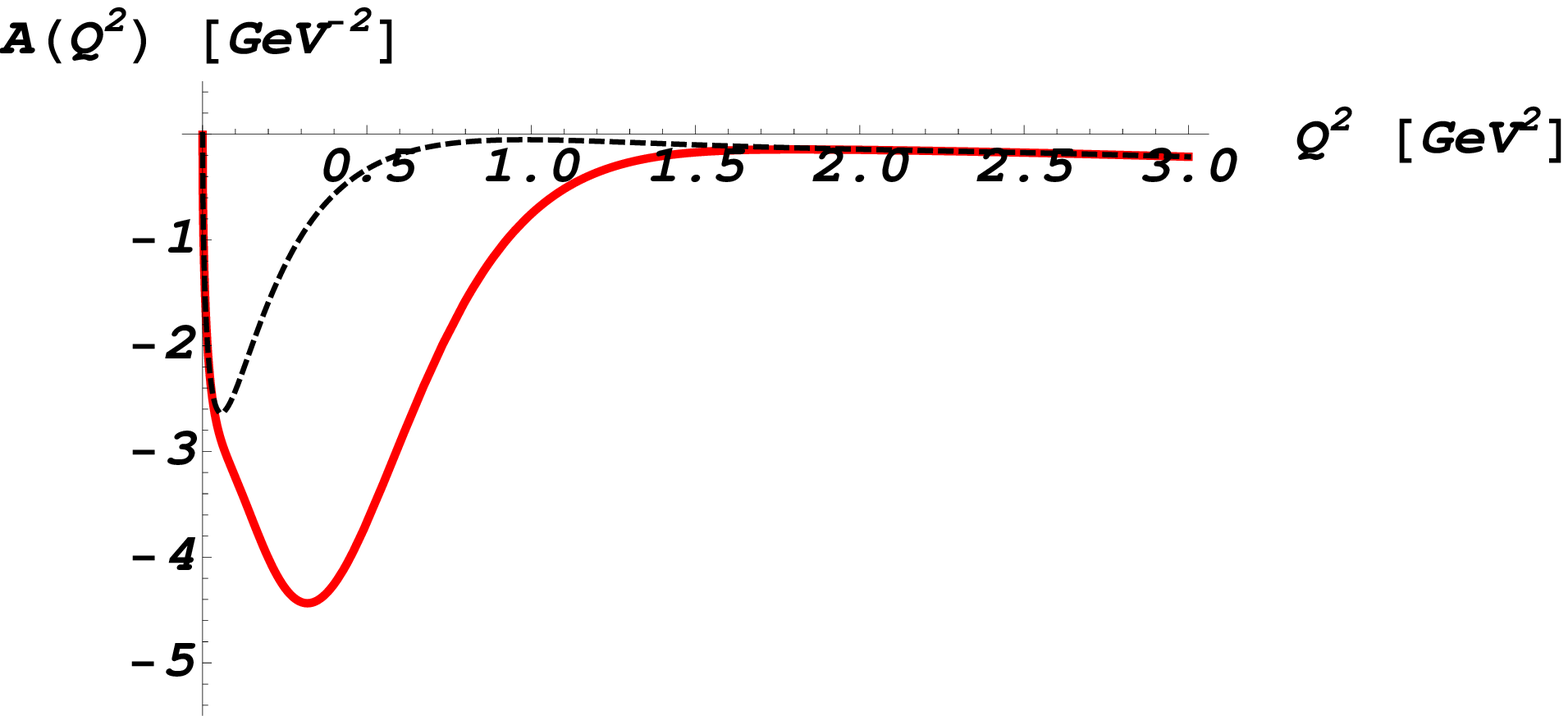}
\includegraphics[width = 6.25cm]{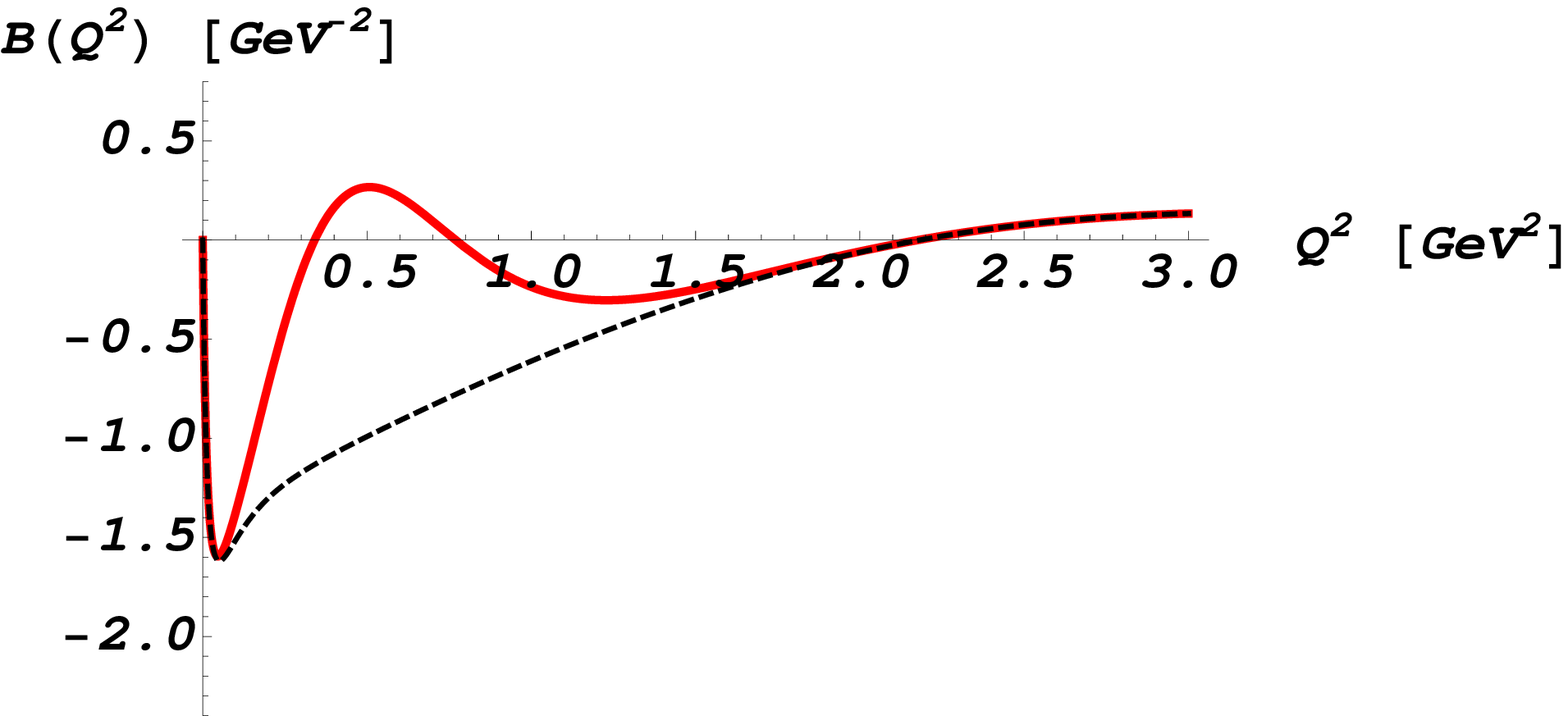}
\includegraphics[width = 6.25cm]{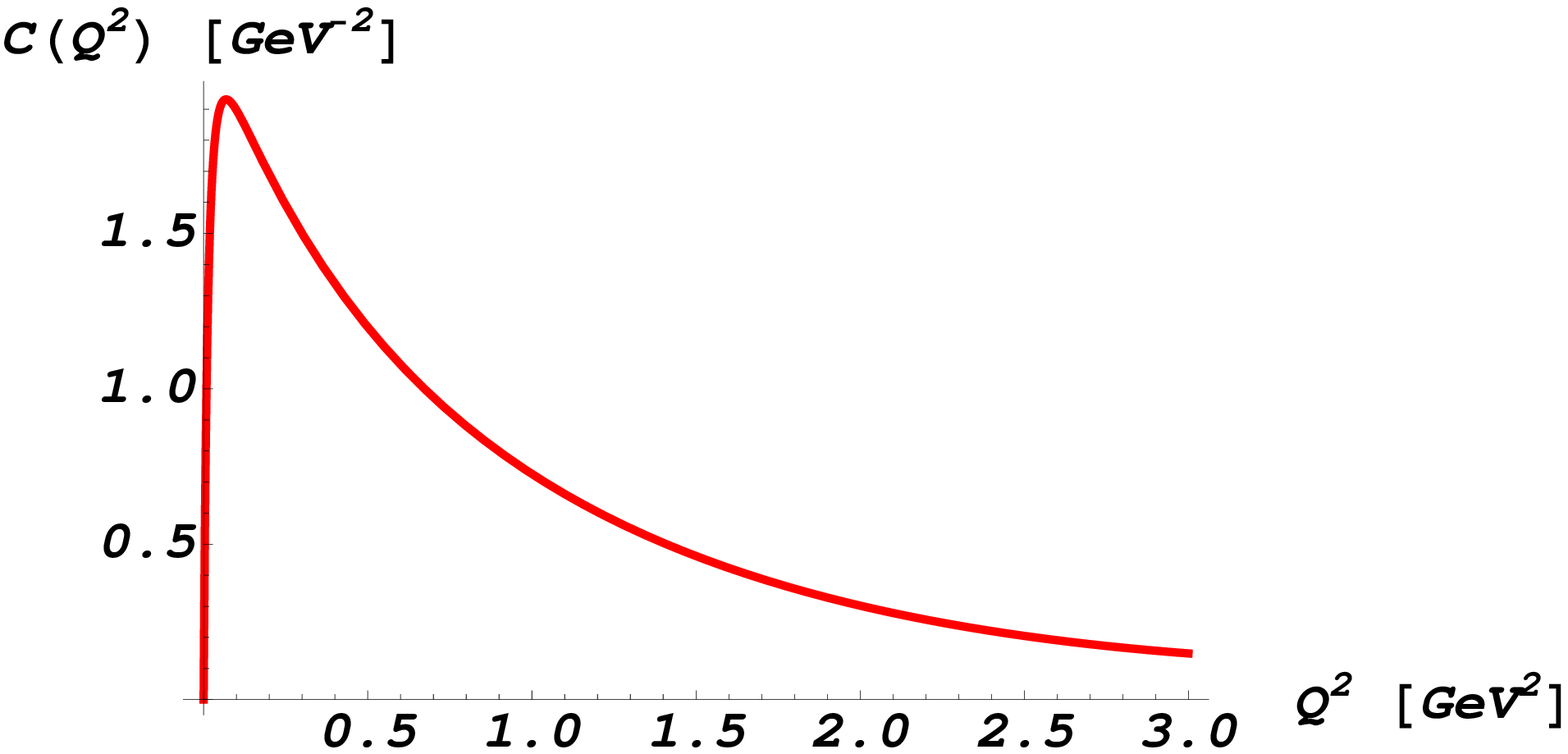}
\end{center}
\vspace{-0.25cm}
\caption{$Q^2$ dependence of the functions A, B, and C.  
Dotted (black) curves are for parameterization GP I; 
solid (red) curves are for GP II,  see caption of Fig.~\ref{fig:pllplt}. 
}
\label{fig:abc}
\end{figure}

To evaluate the induced polarizations, we use the available empirical information  on the GPs. 
The four spin GPs are described, following~\cite{Pasquini:2001yy},  
by a dispersive part, and a  $\pi^0$ pole part. 
The dispersive part is saturared by $\pi N$ intermediate states, using 
empirical information from pion photo- and electroproduction as encoded in the MAID2007 parameterization~\cite{Drechsel:2007if}. 
The electric and magnetic GPs are  decomposed as a sum of a dispersive $\pi N$ part and an asymptotic part. The asymptotic part of the magnetic GP is described by a dipole~: 
\begin{eqnarray}
P^{(M1,M1)0}_{asy}(Q^2) &=& P^{(M1,M1)0}_{asy}(0) / (1 + Q^2 / \Lambda_\beta^2)^2 .  \;\;\;\;\;
\label{eq:gpmagn}
\end{eqnarray}
To describe the available data for the electric GP, we allow for an asymptotic part consisting of 
a sum of a dipole and a gaussian, in the same vein as the parameterization 
proposed in \cite{Friedrich:2003iz} for the nucleon form factors~:
\begin{eqnarray}
P^{(L1,L1)0}_{asy}(Q^2) &=& P^{(L1,L1)0}_{asy}(0) / (1 + Q^2 / \Lambda_\alpha^2)^2 
\nonumber \\
&+& C_\alpha \, Q^4 \, e^{- (Q^2 - 0.15) / 0.15}.
\label{eq:gpelec} 
\end{eqnarray}
The values at the real photon point have been fixed as the difference between the empirical information 
for the proton electric and magnetic polarizabilities, obtained from real Compton scattering (RCS) 
experiments~\cite{OlmosdeLeon:2001zn}, and the dispersive $\pi N$ contribution, yielding
$P^{(L1,L1)0}_{asy}(0) = -14.37$~GeV$^{-3}$, and 
$P^{(M1,M1)0}_{asy}(0) = 21.82$~GeV$^{-3}$. 
The remaining three parameters $\Lambda_\beta, \Lambda_\alpha$, and $C_\alpha$, 
describing the $Q^2$ dependence of the asymptotic parts of the spin independent GPs can be determined by a fit to available VCS data.  
In Fig.~\ref{fig:pllplt}, we show the comparison with the experimentally measured unpolarized structure functions $P_{LL} - P_{TT} / \varepsilon$ ($P_{LT}$), proportional to the electric (magnetic) GPs respectively, up to a small spin GP contribution (dashed curves). 
For an exhaustive description of VCS observables, we refer to 
Refs.~\cite{Guichon:1998xv, Drechsel:2002ar}. 
For the magnetic GP, one sees from $P_{LT}$ on Fig.~\ref{fig:pllplt} that 
a good fit to all data is obtained for $\Lambda_\beta = 0.5$~GeV. 
For the electric GP, a fit to the MIT-Bates and JLab data is obtained for $C_\alpha = 0$, 
and $\Lambda_\alpha = 0.7$~GeV (denoted by parameterization GP I). 
However, this does not describe the MAMI data 
at intermediate $Q^2$, which require an additional structure, parameterized through 
the gaussian term in Eq.~(\ref{eq:gpelec}). A good description of all available data is obtained for 
$\Lambda_\alpha = 0.7$~GeV, and $C_\alpha = -150 \;  \mathrm{GeV}^{-7}$ 
(denoted by parameterization GP II). 
The above empirical parameterizations for the GPs, allow to evaluate the 
$Q^2$ dependence of $A, B$, and $C$, as displayed in Fig.~\ref{fig:abc}. 
One clearly notices the enhancement in $A$ as well as the structure at intermediate $Q^2$ values 
in $B$ in GP II. 

The light-front frame allows us to use this empirical information to 
visualize the deformation of the charge densities in an 
external e.m. field and map out the  transverse position space dependence of the 
induced polarization. 
For the case of a nucleon in a state of definite helicity, the transverse position space dependence 
of the induced polarization $\vec P_0$ is given by~:
\begin{equation}
\vec P_0(\vec b) = \int \frac{d^2 \vec q_\perp}{(2 \pi)^2} \, e^{- i \vec q_\perp \cdot \vec b} \, 
\vec P_0(\vec q_\perp),
\end{equation}
which can be worked out using Eq.~(\ref{eq:indpolnoflip}) as
\begin{equation}
\vec P_0(\vec b) = \hat b \, \int_0^\infty \frac{d Q}{(2 \pi)} \, Q  \, J_1(b \, Q) \,  A(Q^2),
\label{eq:indpolnoflipb}
\end{equation}
where $\vec b$ is the transverse position, $b = | \vec b|$, and $\hat b = \vec b / b$.  

\begin{figure}
\begin{center}
\includegraphics[width = 6.25cm]{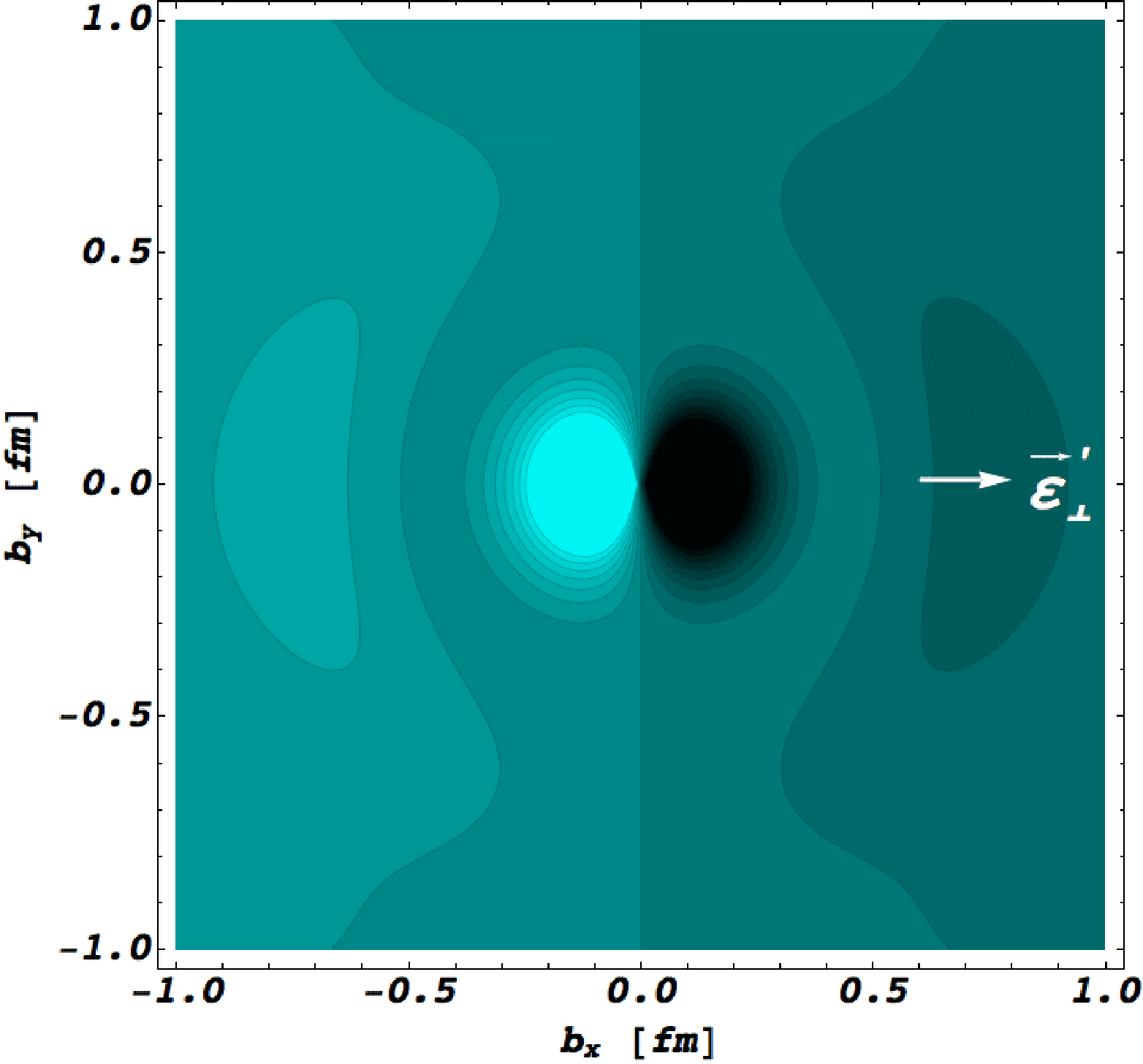}
\includegraphics[width = 6.25cm]{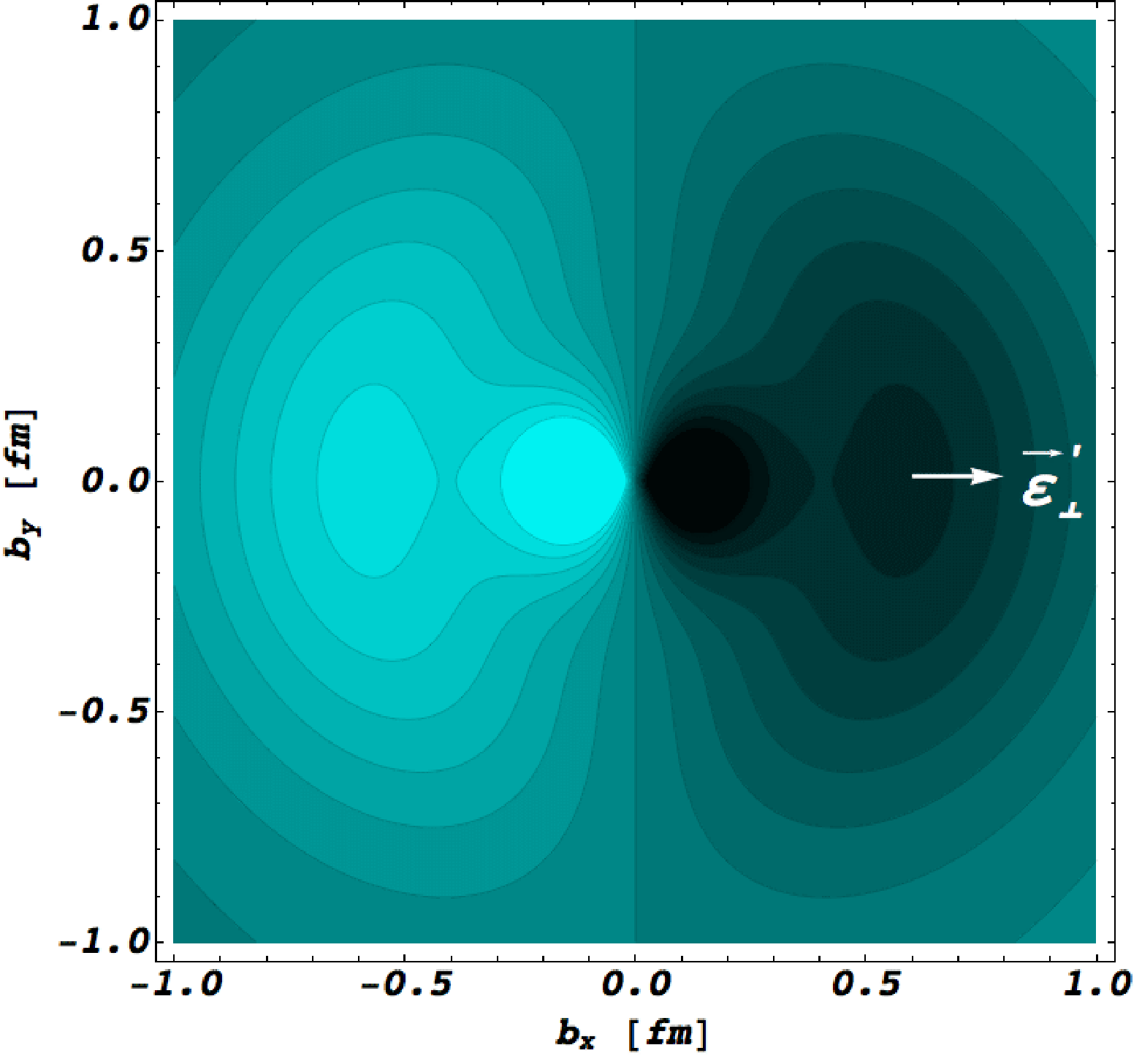}
\end{center}
\begin{center}
\includegraphics[width = 7.25cm]{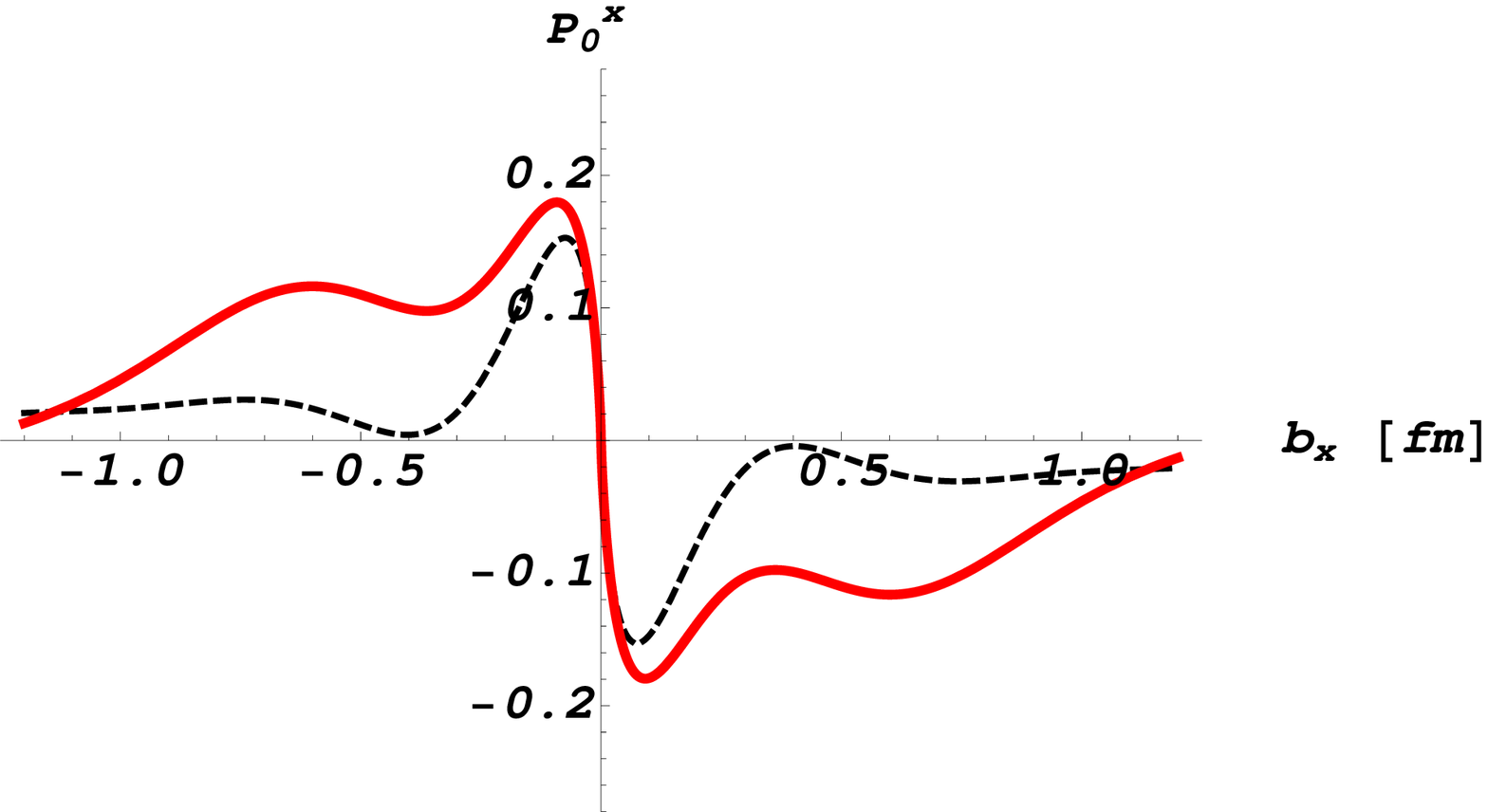}
\end{center}
\vspace{-0.35cm}
\caption{Induced polarization, $P_0^x$, in a proton of 
definite light-cone helicity, when submitted to an e.m. field with photon polarization along the $x$-axis, 
as indicated. 
Upper (middle) panel is for GP~I (GP II), see caption of Fig.~\ref{fig:pllplt}.
The light (dark) regions correspond to the
largest (smallest) values.  
The lower panel compares $P_0^x$ along $b_y = 0$~: dotted curve is for GP I; solid curve is for GP II.   
}
\label{fig:p0x}
\end{figure}

The dipole pattern described by Eq.~(\ref{eq:indpolnoflipb}) is shown in Fig.~\ref{fig:p0x}. 
One clearly sees that the enhancement at intermediate $Q^2$ in the electric GP (upper panel in 
Fig.~\ref{fig:pllplt}) in GP II, as compared with GP I, yields a spatial distribution of 
the induced polarization that extends noticeably to larger transverse distances. Forthcoming 
VCS experiments, that are conceived to pin down more precisely the behavior of the GPs at intermediate $Q^2$ values, will therefore  be able to verify this large distance structure.   

\begin{figure}
\begin{center}
\includegraphics[width = 6.25cm]{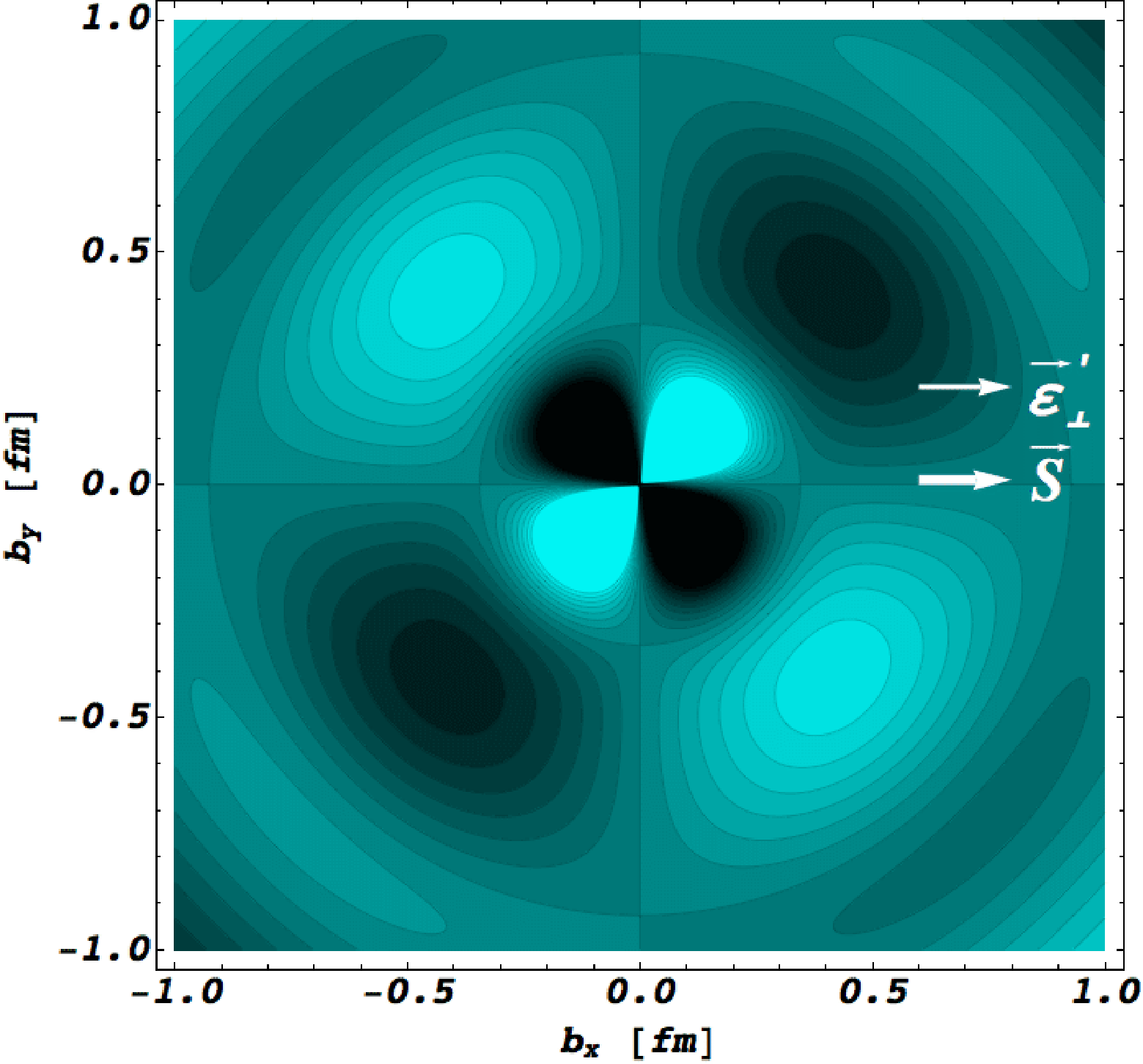}
\includegraphics[width = 6.25cm]{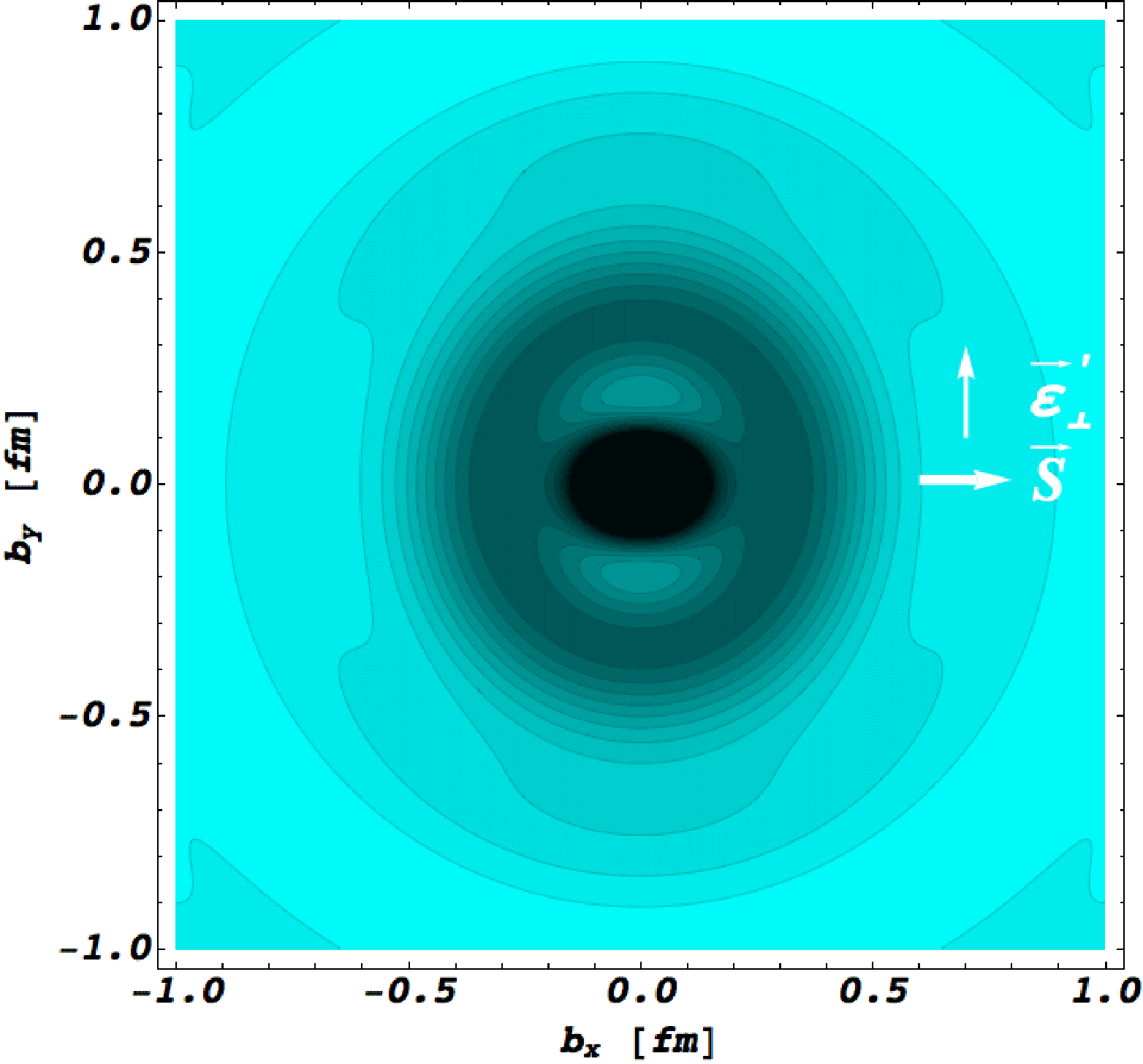}
\end{center}
\vspace{-0.35cm}
\caption{Induced polarization density in a proton, with spin $\vec S$ oriented along the $x$-axis, when submitted to an e.m. field.  
The upper (lower) panels are for $P_T^x - P_0^x$ ($P_T^y - P_0^y$) respectively, 
and correspond with photon polarization along the $x$-axis ($y$-axis) as indicated. 
The light (dark) regions correspond to the
largest (smallest) values using parameterization GP II.  
}
\label{fig:ptxy}
\end{figure}

For the case of a nucleon in an eigenstate of transverse spin, the transverse position 
dependence of the induced polarization $\vec P_T$ can likewise be worked out as~:
\begin{eqnarray}
\vec P_T (\vec b) &=&  \vec P_0(\vec b) \\
&-&  \hat b \, \left( \vec S_\perp \times \vec e_z \right) \cdot \hat b \, 
\int_0^\infty \frac{d Q}{(2 \pi)} \, Q  \, J_2(b Q) \,  B \nonumber \\
&+&  \left( \vec S_\perp \times \vec e_z \right)   
\int_0^\infty \frac{d Q}{(2 \pi)} Q  \left[ J_0(b Q)   C + \frac{J_1(b Q)}{bQ}  B 
\right], \nonumber 
\end{eqnarray}
displaying dipole, quadrupole and monopole patterns. 
In Fig.~\ref{fig:ptxy} we show the spatial distributions in the induced polarization 
for a proton of transverse spin (chosen along the $x$-axis) for parameterization GP II.  The component  $P_T^x - P_0^x$ displays a quadrupole pattern with pronounced strength around $0.5$~fm due to the electric GP, whereas the component $P_T^y - P_0^y$ shows in addition a monopole pattern, dominated by the $\pi^0$ pole contribution. 

In summary, in this work we have used recent data on proton GPs
to map out the spatial dependence of the induced polarizations in an external e.m. field. 
The formalism to extract in a field theoretic consistent way 
light-front densities from nucleon form factor data 
has been extended in this work to the deformations of these quark charge densities when 
applying an external e.m. field. It has been shown that the available proton electric GP data yield a pronounced structure in its induced polarization at large transverse distances of $0.5 - 1$~fm. 
At $Q^2$ values smaller than 0.1 GeV$^2$, chiral effective field theory was found to well describe the VCS data, highlighting the role of pions in the nucleon structure. Such description can however not be applied at intermediate and large $Q^2$ values. This transition region is dominated by nucleon resonance structure, which can be described by dispersion relations. 
Forthcoming VCS precision experiments at MAMI in this intermediate $Q^2$ region, will be able to better determine this structure, thus complementing our picture of the distribution of quark charges in the nucleon as obtained through elastic form factors. 


\end{document}